\documentclass{elsart}

\usepackage{cite}
\usepackage{graphicx}

\begin{document}

\begin{frontmatter}

\title{On the application of homotopy perturbation method to
differential equations}

\author{Francisco M. Fern\'{a}ndez \thanksref{FMF}}

\address{INIFTA (UNLP, CCT La Plata-CONICET), Divisi\'{o}n Qu\'{i}mica Te\'{o}rica,\\
Diag. 113 y 64 (S/N), Sucursal 4, Casilla de Correo 16,\\
1900 La Plata, Argentina}

\thanks[FMF]{e--mail: fernande@quimica.unlp.edu.ar}

\begin{abstract}
We show that a recent application of homotopy perturbation method to a class
of ordinary differential equations yields either useless or wrong results.
\end{abstract}

\end{frontmatter}

There has recently been great interest in the application of several
approximate procedures, like the homotopy perturbation method (HPM), the
Adomian decomposition method (ADM), and the variation iteration method
(VIM), to a variety of linear and nonlinear problems of interest in
theoretical physics \cite
{RDGP07,CHA07,YO07,CH07a,EG07,GAHT07,SNH07,CH08,ZLL08,SNH08b,M08,SNH08,RAH08,YE08,SG08}%
. For brevity I will call VAPA all those variational and perturbational
approaches. In a series of papers I have shown that most of the results
produced by those methods are useless, nonsensical, and worthless\cite
{F07,F08b,F08c,F08d}.

In a recent paper Rafiq et al\cite{RAH08} applied the HPM to some ordinary
second--order differential equations, and the purpose of this comment is to
discuss their results.

Rafiq et al\cite{RAH08} solved second order differential equations of the
form
\begin{equation}
\left\{
\begin{array}{l}
y^{\prime \prime }(x)+p(x)y^{\prime }(x)+f(x,y)=0 \\
y(0)=A,\;y^{\prime }(0)=B
\end{array}
\right.  \label{eq:diff}
\end{equation}
where $f(x,y)=F(x,y)-g(x)$ in their notation. In order to apply HPM they
wrote $y^{\prime \prime }(x)+p(x)y^{\prime }(x)+\theta f(x,y)=0$ and
expanded the solution in a $\theta $--power series
\begin{equation}
y(x)=y_{0}(x)+\theta y_{1}(x)+\ldots  \label{eq:HPMseries}
\end{equation}
Finally, they set $\theta =1$ in order to obtain an approximate solution to
equation (\ref{eq:diff}).

In the particular HPM implementation proposed by Rafiq et al\cite{RAH08} the
perturbation corrections $y_{j}(x)$ result to be polynomials so that the
partial sums of the HPM series (\ref{eq:HPMseries}) are just $x$--power
series of the form
\begin{equation}
y(x)=A+Bx+a_{2}x^{2}+a_{3}x^{3}+\ldots  \label{eq:y_x_series}
\end{equation}

In what follows we analyze the examples chosen by Rafiq et al\cite{RAH08}.
The first one is a textbook exercise for an introductory course on
differential equations:
\begin{equation}
\left\{
\begin{array}{l}
y^{\prime \prime }(x)+\frac{8}{x}y^{\prime }(x)+xy=x^{5}+44x^{2}-30x \\
A=B=0
\end{array}
\right.   \label{eq:Ex1}
\end{equation}
We appreciate that $y(x)\sim x^{4}$ as $x\rightarrow \infty $. Therefore, if
we substitute the polynomial $y(x)=a_{2}x^{2}+a_{3}x^{3}+a_{4}x^{4}$ into
equation (\ref{eq:Ex1}) we easily obtain the exact solution $y(x)=x^{4}-x^{3}
$.  However, Rafiq et al\cite{RAH08} apply a cumbersome perturbation method
and obtain an infinite number of nonzero perturbation corrections $y_{j}(x)$
that cancel out to (hopefully) produce the exact solution. The authors do
not prove that their HPM already yields the exact result, and any partial
sum gives the exact result plus terms that are cancelled by corrections of
higher order. In this case the HPM partial sum of any order is an inexact
approach to an extremely simple solution of a trivial differential equation
that one easily derives by the straightforward method just indicated.

The second example is as trivial as the first one:
\begin{equation}
\left\{
\begin{array}{l}
y^{\prime \prime }(x)+\frac{2}{x}y^{\prime }(x)+y=6+12x+x^{2}+x^{3} \\
A=B=0
\end{array}
\right.  \label{eq:Ex2}
\end{equation}
Any undergraduate student will try $y(x)=a_{2}x^{2}+a_{3}x^{3}$ and obtain
the exact solution $y(x)=x^{2}+x^{3}$ without effort. Again, the HPM yields
an infinite series and the authors do not prove their convergence. Although
it seems that the spurious terms cancel out, any partial sum yields a wrong
result.

A simple inspection of the third example
\begin{equation}
\left\{
\begin{array}{l}
y^{\prime \prime }(x)+\frac{2}{x}y^{\prime }(x)+y^{3}=6+x^{6} \\
A=B=0
\end{array}
\right.  \label{eq:Ex3}
\end{equation}
suggests that $y(x)=a_{2}x^{2}$ and thus one obtains the exact result $%
y(x)=x^{2}$. By means of the HPM Rafiq et al\cite{RAH08} obtain the wrong
result $y(x)=x^{2}+x^{8}/72$!!! Chowdhury and Hashim\cite{CH07a} also obtain
this wrong result but then they choose a different starting point of the
perturbation approach in order to derive the expected exact solution. In the
first two examples Rafiq et al\cite{RAH08} mention the appearance of noise
terms. It seems to me that it was that noise that already affected the
calculation in this example.

The fourth example is much more interesting (at least it is not trivial):
\begin{equation}
\left\{
\begin{array}{l}
y^{\prime \prime }(x)+\frac{2}{x}y^{\prime }(x)+e^{xy^{2}}=x+1 \\
A=B=0
\end{array}
\right.   \label{eq:Ex4}
\end{equation}
After a long and tedious perturbation calculation the authors derive the
first correct terms of the Taylor series expansion about $x=0$\cite{RAH08}
\begin{equation}
y(x)=\frac{x^{3}}{12}-\frac{x^{9}}{12960}+\ldots   \label{eq:Ex4_y_series}
\end{equation}
It is unnecessary to say that one can easily obtain the same approximate
result more easily by means of the power--series method. We expect that the
power series will not reveal the most interesting features of the solution
to this equation but a mere indication of what happens in a neighbourhood of
$x=0$.

\begin{figure}[H]
\begin{center}
\includegraphics[width=9cm]{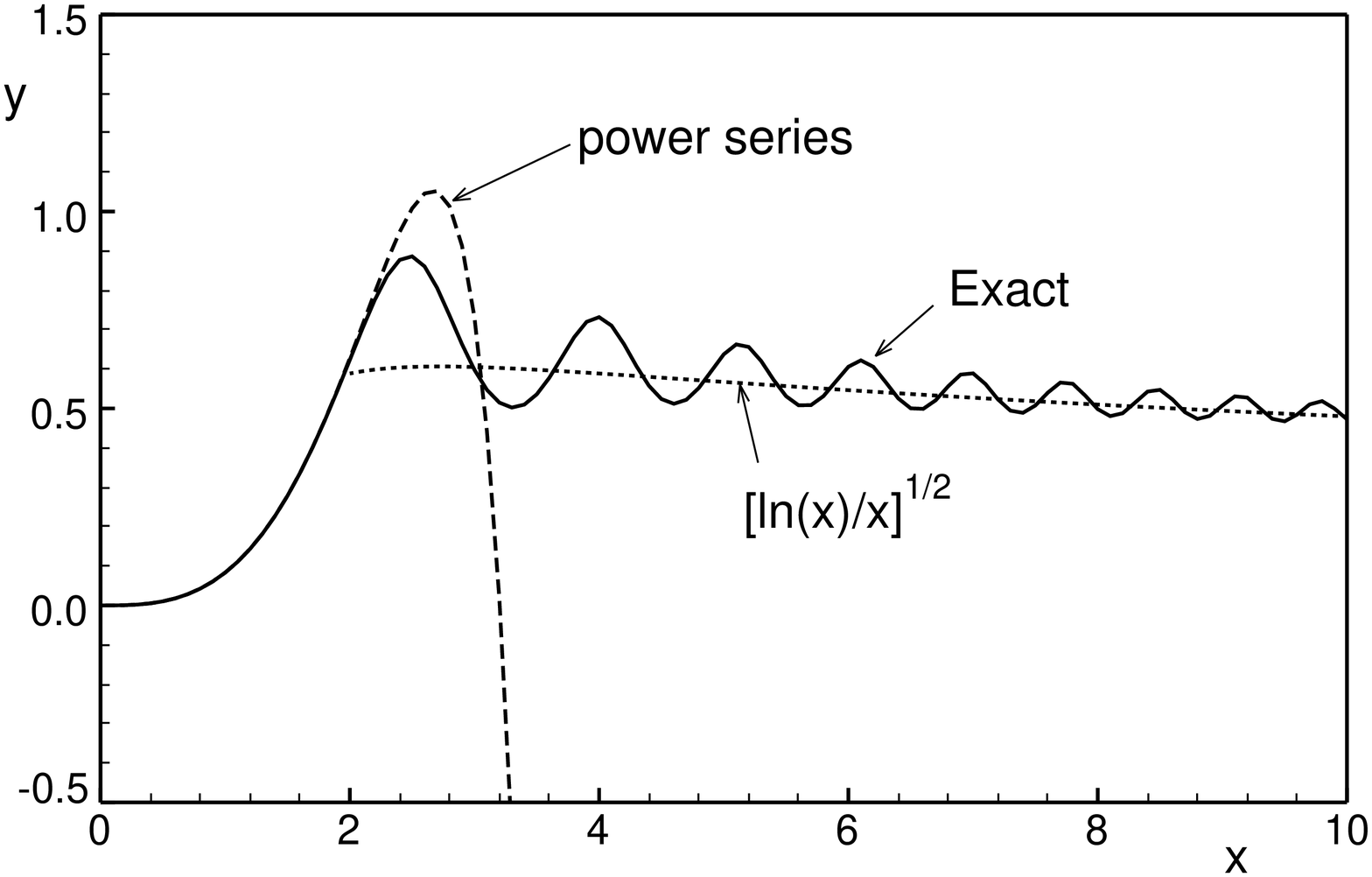}
\end{center}
\caption{Exact (accurate numerical) solution, power series~(\ref
{eq:Ex4_y_series}) and asymptotic expansion for example~(\ref{eq:Ex4})}
\label{fig:ex4}
\end{figure}

At first sight one guesses that the solution to equation (\ref{eq:Ex4})
should behave approximately as $y(x)\sim \sqrt{\ln (x)/x}$ as $x\rightarrow
\infty $; however, its actual behaviour is richer. Fig.~\ref{fig:ex4} shows
that $y(x)$ (calculated numerically with sufficient accuracy) oscillates
about $\sqrt{\ln (x)/x}$ and approaches this asymptotic function as $%
x\rightarrow \infty $. The $x$--power series (\ref{eq:Ex4_y_series}) only
accounts for the behaviour of $y(x)$ up to about the first maximum. If
equation (\ref{eq:Ex4}) represented an actual physical problem we would be
missing its most interesting features when using the HPM to obtain its
solution.

Summarizing: the HPM proposed by Rafiq et al\cite{RAH08} gives cumbersome
approximations to simple solutions of trivial differential equations, yields
a wrong result in one of the cases, and fails to provide the most
interesting features of the solution to the only nontrivial example.
Unfortunately the authors do not indicate any physical application of the
examples chosen. They seem to be just toy problems for fiddling around with
the HPM.

The conclusion above is not surprising because all our previous analysis of
several applications of VAPA have shown that they produce useless or trivial
results\cite{F07,F08b,F08c,F08d}. See, for example, the earlier papers by
Chowdhury and Hashim\cite{CH07a} and Y\i ld\i r\i m and \"{O}zi\c{s}\cite
{YO07} who applied similar cumbersome perturbation equations and obtained
the power series for the solutions of exactly solvable differential
equations.

It seems that one of the greatest feats of many VAPA applications is to
produce power--series expansions to simple and trivial problems in a
cumbersome and laborious way\cite{RAH08,CH07a,YO07}; the reader may find the
analysis of other such examples in our earlier communications\cite
{F07,F08b,F08c,F08d}. In fact, VAPA have produced the worst research papers
ever written.

\end{document}